\documentclass[]{aa}

\usepackage{graphicx}
\usepackage{txfonts}

\newcommand{\be}{\begin{equation}}
\newcommand{\ee}{\end{equation}}

\begin{document}

\title{Hadronic gamma-ray emission from windy microquasars}

\author{G. E. Romero\inst{1,}\thanks{Member of CONICET}, D. F.
Torres\inst{2}, M. M. Kaufman Bernad\'o,\inst{1} and
I.F. Mirabel\inst{3,4,}$^{\star}$}

\offprints{G.E. Romero\\ \email{romero@venus.fisica.unlp.edu.ar}}

\institute{Instituto Argentino de Radioastronom\'{\i}a, C.C.5,
(1894) Villa Elisa, Buenos Aires, Argentina \and Lawrence
Livermore National Laboratory, 7000 East Avenue, L-413, Livermore,
CA 94550, USA \and CEA/DSM/DAPNIA/Service d'Astrophysique, Centre
d'Etudes de Saclay, F-91191 Gif-sur-Yvette, France \and Instituto
de Astronom\'{\i}a y F\'{\i}sica del Espacio/CONICET, C.C. 67,
Suc. 28, Buenos Aires, Argentina}

\date{Received / Accepted}

\abstract{The jets of microquasars with high-mass stellar
companions are exposed to the dense matter field of the stellar
wind. We present estimates of the gamma-ray emission expected from
the jet-wind hadronic interaction and we discuss the detectability
of the phenomenon at high energies. The proposed mechanism could
explain some of the unidentified gamma-ray sources detected by
EGRET instrument on the galactic plane. \keywords{X-ray binaries
-- stars -- gamma-rays: observations -- gamma-rays: theory}}

\titlerunning{Gamma-rays from microquasars}

\authorrunning{G.E. Romero et al.}

\maketitle

\section{Introduction}

Microquasars (MQs) are X-ray binary systems that present non-thermal
jet-like features (Mirabel \& Rodr\'{\i}guez 1999).
The compact jets can be detected at radio wavelengths with flat spectra.
The presence of apparent superluminal movements in some cases (e.g.
Mirabel \& Rodr\'{\i}guez 1994) indicates the existence of
relativistic bulk motions in the jet flow. Very recently,
Chandra observations have revealed X-ray synchrotron emission from the jet
of XTE J1550-564 (Corbel et al. 2002, Kaaret et al. 2003), a fact
that indicates the presence of extremely relativistic electrons
with TeV energies and shock reacceleration in the jet.
The interaction of such highly energetic particles with
the ambient photon fields will produce high-energy inverse Compton
(IC) radiation. In a series of recent papers,
Aharonian \& Atoyan
(1998), Atoyan \& Aharonian (1999), Markoff et al. (2001),
Georganopoulos et al. (2002), and Romero et al. (2002) have
studied different aspects of the synchrotron and the IC emission
from MQs whose jets interact with different photon
fields. 
Recently, Paredes et al.
(2000) suggested that the MQ LS 5039 might be responsible
for the gamma-ray source 3EG J1824-1514. Kaufman Bernad\'o et al.
(2002) proposed that high-mass MQs with jets forming a
small viewing angle (i.e. {\sl microblazars}) might be the parent
population of the set of bright and variable gamma-ray sources
detected by EGRET along the galactic plane (e.g. Romero
et al. 1999, Torres et al. 2001). The main gamma-ray
production mechanism for MQs advocated by Kaufman
Bernad\'o et al. (2002) was IC upscattering of UV photons from the
stellar companion. However, pair creation absorption processes in
the disk and coronal X-ray field might quench the source above a
few MeV in many cases (Romero et al.  2002).

In this {\em Letter} we present a new mechanism for the generation
of high-energy gamma-rays in MQs that is based on
hadronic interactions occurring outside the coronal region.
The gamma-ray emission arises from the decay of neutral pions
created in the inelastic collisions between relativistic protons ejected by
the compact object and the ions in the stellar wind. The only requisites for the model are a windy high-mass stellar
companion and the presence of multi-TeV protons in the
jet\footnote{Interactions
of hadronic beams with moving clouds in
the context of accreting pulsars have been previously discussed in
the literature by Aharonian \& Atoyan (1996). For an early
discussion in a general context see Bednarek et al (1990).}. The
presence of relativistic hadrons in MQ jets like those of
SS 433 has been inferred from iron X-ray line observations (e.g.
Kotani et al. 1994, 1996; Migliari et al. 2002), although direct and clear
evidence exists only for this source so far.
In what follows we describe the model and
present the results of our calculations.

\section{The jet}

\begin{figure}
\resizebox{6.5cm}{!}{\includegraphics{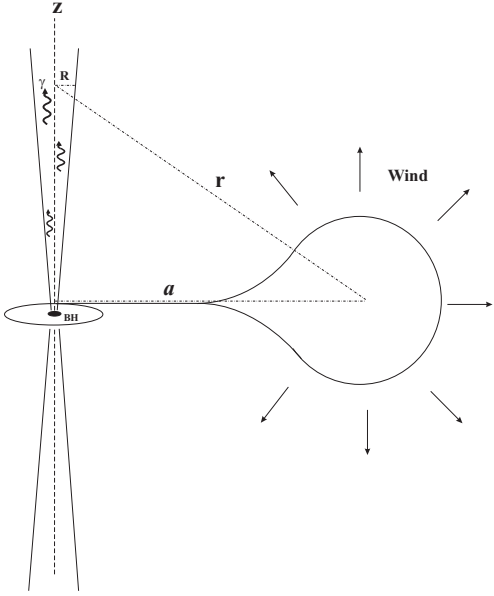}} \caption{\rm
Sketch of the general situation discussed in the paper. A
relativistic $e-p$ jet is injected close to the black hole in a
MQ with a high-mass stellar companion. The jet must
traverse the matter field created by the stellar wind. The
resulting interaction produces gamma-ray emission. In
the figure perpendicularity is assumed between the jet and the orbital
plane, but this particular assumption can be relaxed in a more general situation.} \label{fig1}
\end{figure}

The general situation discussed in this paper is shown in
Fig.~\ref{fig1}. A binary system is formed by a black hole and a
high-mass early-type star. A relativistic $e-p$ jet is ejected by
the compact object perpendicularly to the accretion disk plane.
For simplicity, we shall assume that this is also the orbital
plane, but this condition can be relaxed to allow, for instance, a
precessional motion (Kaufman Bernad\'o et al. 2002) or more
general situations as discussed by Maccarone (2002) and Butt et al. (2003).


The jet axis, $z$, is assumed to be normal to the orbital radius $a$. We shall
allow the jet to expand laterally, in such a way that its radius
is given by $R(z)=\xi z^{\epsilon}$, with $\epsilon\leq 1$ and
$z_0\leq z\leq z_{\rm max}$. For $\epsilon=1$ we have a conical
beam. The jet starts to expand at a height $z_0\sim$ a few hundred
km above the black hole, outside the coronal region. The particle
spectrum of the relativistic $e-p$ flow is assumed to be a power
law $N'_{e,\;p}(E'_{e, \;p})= K_{e, \;p}\; {E'}_{e,
\;p}^{-\alpha}$, valid for $ {E'_{e, \;p}}^{\rm min}\leq  E'_{e,
\;p} \leq  {E'_{e, \;p}}^{\rm max}$, in the jet frame. The
corresponding particle flux will be $J'_{e, \;p}( E'_{e, \;p})=
(c/4\pi) N'_{e,\;p}(E'_{e,\;p})$. Since the jet expands, the
proton flux can be written as:
\begin{equation}
J'_p(E'_p)=\frac{c}{4 \pi} K_0
\left(\frac{z_0}{z}\right)^{\epsilon n} {E'_p}^{-\alpha},
\label{Jp}
\end{equation}
where $n>0$ (a value $n=2$ corresponds to the conservation of the
number of particles, see Ghisellini et al. 1985), and a prime
refers to the jet frame. Using relativistic invariants, it can be
proven that the proton flux, in the observer (or lab) frame,
becomes (e.g. Purmohammmad \& Samimi 2001)
\begin{equation}
J_p(E_p,\theta)=\frac{c K_0}{4 \pi} \left(\frac{z_0}{z}\right)^
{\epsilon n} \frac{\Gamma^{-\alpha+1} \left(E_p-\beta_{\rm b}
\sqrt{E_p^2-m_p^2c^4} \cos \theta\right)^{-\alpha}}{\left[\sin ^2
\theta + \Gamma^2 \left( \cos \theta - \frac{\beta_{\rm b}
E_p}{\sqrt{E_p^2-m_p^2 c^4}}\right)\right]^{1/2}}, \label{Jp_lab}
\end{equation}
where $\Gamma$ is the jet Lorentz factor, $\theta$ is the angle
subtended by the emerging photon direction and the jet axis, and $\beta_{\rm b}$
is the corresponding velocity in units of $c$.
The exponential dependence of the cross
section on the transverse
momentum ($p_{\rm t}$) of the incident protons beams the
gamma-ray emission into an angle $\phi<cp_{\rm t}/m_{\rm p} \Gamma\sim 0.17/\Gamma$
along the proton direction, hence justifying the assumption that both directions are similar.
Note that only
photons emitted with angles similar to that of the inclination
angle of the jet will reach a distant observer, and thus $\theta$
can be approximated by the jet inclination angle.

In order to determine the matter content of the jet we will adopt
the jet-disk coupling hypothesis proposed by Falcke \& Biermann
(1995) and applied with success to AGNs (see also Falcke \&
Biermann 1996), i.e. the total jet power scales with the accreting
rate as $Q_{\rm j}=q_{\rm j} \dot{M}_{\rm disk} c^2$, with $q_{\rm
j}=10^{-1}-10^{-3}$. The number density ${n_0}'$ of particles
flowing in the jet at $R_0=R(z_0)$ is then given by $c\pi R_0^2
{n_0}'=Q_{\rm j}/m_p c^2$, where $m_p$ is the proton rest mass.
From here we can obtain ${n_0}'$.
Additionally, $n_0'=\int^{{E'}_p^{\rm max}}_{{E'}_p^{\rm min}}
{N'}_p({E'}_p, \;z_0)\; d{E_p}'$.
Then, if ${E'}_p^{\rm max}>>{E'}_p^{\rm min}$, which is always the
case, we have
$K_0={n_0}' (\alpha-1) ({E'}_p^{\rm min})^{\alpha-1},\label{n02}$
which gives the constant in the power-law spectrum at $z_0$.

\section{The wind}

Early-type stars, like OB stars, lose a significant fraction of
their masses through very strong supersonic winds. Typical mass
loss rates and terminal wind velocities for O stars are of the
order of $10^{-5}$ $\dot{M_{\sun}}$ yr$^{-1}$ and 2500 km
s$^{-1}$, respectively (Lamers \& Cassinelli 1999). At the base of
the wind, the density can easily reach $10^{-12}$ g cm$^{-3}$.
Such strong winds provide a field of matter dense enough as to
produce significant hadronic gamma-rays when pervaded by a
relativistic beam.

The structure of the matter field will be determined essentially
by the stellar mass loss rate and the continuity equation:
$\dot{M_*}=4\pi r^2 \rho (r) v(r)$, where $\rho$ is the density of
the wind and $v$ is its velocity.
The radial dependence of the wind velocity is given by
(Lamers \& Cassinelli 1999):
\begin{equation}
v(r)=v_{\infty}\left(1-\frac{r_*}{r}\right)^{\beta},
\end{equation}
where $v_{\infty}$ is the terminal wind velocity, $r_*$ is the
stellar radius, and the parameter $\beta$ is $\sim 1$ for massive
stars. Hence, using the fact that $r^2=z^2+a^2$ and assuming a gas
dominated by protons, we get the particle density of the medium
along the jet axis:
\begin{equation} n(z)=\frac{\dot{M_*}}{4\pi m_p v_{\infty}
(z^2+a^2)}\left(1-\frac{r_*}{\sqrt{z^2+a^2}}\right)^{-\beta}.\label{n(z)}
\end{equation}

The stability of a relativistic jet under the effects of an
external wind has been recently investigated by Hardee \& Hughes
(2003) through both theoretical analysis and numerical
simulations. Their results indicate that jets surrounded by
outflowing winds are in general more dynamically stable than those
surrounded by a stationary medium. Note that protons pertaining to
the wind can diffuse into the jet medium. The wind penetration into the jet outflow
depends on the parameter $\varpi \sim v R(z)/D$,
where $v$ is velocity of wind, $R(z)$ is
the radius of the jet at a height $z$ above the compact object,
and $D$ is the diffusion coefficient. $\varpi$ measures the ratio
between the diffusive and the convective timescale of the
particles. In the Bohm limit, with typical magnetic fields $B_0\sim 1-10$ G,
$\varpi \leq 1$, and the wind matter penetrates the jet by diffusion.

\section{Gamma-ray emission}


Pion decay chains leads to gamma-ray and neutrino production.
The differential
gamma-ray emissivity from $\pi^0$-decays is:
\begin{equation}
q_{\gamma}(E_{\gamma})= 4 \pi \sigma_{pp}(E_p)
\frac{2Z^{(\alpha)}_{p\rightarrow\pi^0}}{\alpha}\;J_p(E_{\gamma},\theta)
\eta_{\rm A}. \label{q} \end{equation} Here, the parameter
$\eta_{\rm A}$ takes into account the contribution from different
nuclei in the wind and in the jet (for standard composition of
cosmic rays and interstellar medium  $\eta_{\rm A}=1.4-1.5$,
Dermer 1986). $J_p(E_{\gamma})$ is the proton flux distribution
evaluated at $E=E_{\gamma}$. The cross section $\sigma_{pp}(E_p)$
for inelastic $p-p$ interactions at energy $E_p\approx 10
E_{\gamma}$ can be represented above $E_p\approx 10$ GeV by
$\sigma_{pp}(E_p)\approx 30 \times [0.95 + 0.06 \log (E_p/{\rm
GeV})]$ mb.  Finally, $Z^{(\alpha)}_{p\rightarrow\pi^0}$ is the
so-called spectrum-weighted moment of the inclusive cross-section.
Its value for different spectral indices $\alpha$ is given, for
instance, in Table A1 of Drury et al. (1994). Notice that
$q_{\gamma}$ is expressed in ph s$^{-1}$ erg$^{-1}$ when we adopt
CGS units.

The spectral gamma-ray intensity (photons per unit of time per
unit of energy-band) is:
\begin{equation}
I_{\gamma}(E_{\gamma},\theta)=\int_V n(\vec{r'})
q_{\gamma}(\vec{r'}) d^3\vec{r'}, \label{I}
\end{equation}
where $V$ is the interaction volume.


Since we are interested here in a general model and not in the
study of a particular source, the spectral energy distribution
$L^{\pi^0}_{\gamma}(E_{\gamma},\theta)=E_{\gamma}^2
I_{\gamma}(E_{\gamma},\theta)$ is a more convenient quantity than
the flux. Using eqs. (\ref{Jp_lab}),
(\ref{n(z)}),  (\ref{q}) and (\ref{I}), we get:
\begin{eqnarray}
&& L^{\pi^0}_{\gamma}(E_{\gamma},\theta)\approx  \;\frac{ q_{\rm
j} z_0^{\epsilon (n-2)} Z^{(\alpha)}_{p\rightarrow\pi^0}}{ 2\pi
m_p^2 v_{\infty}} \;\frac{\alpha-1}{\alpha}\;({E'}_p^{\rm
min})^{\alpha-1} \times \nonumber \\ && \dot{M}_* \dot{M}_{\rm
disk} \; \sigma_{pp}(10\; E_{\gamma})\;
 \frac{\Gamma^{-\alpha+1} \left(E_\gamma-\beta_{\rm b}
\sqrt{E_\gamma^2-m_p^2c^4} \cos
\theta\right)^{-\alpha}}{\left[\sin ^2 \theta + \Gamma^2 \left(
\cos \theta - \frac{\beta_{\rm b} E_\gamma}{\sqrt{E_\gamma^2-m_p^2
c^4}}\right)\right]^{1/2}} \nonumber \\ && \int_{z_0}^{\infty}
\frac{z^{\epsilon (2-n)}}{(z^2+a^2)}
\left(1-\frac{r_*}{\sqrt{z^2+a^2}}\right)^{-\beta} dz.
%
\end{eqnarray}
This expression gives approximately the $\pi^0$-decay gamma-ray
luminosity for a windy MQ at energies $E_{\gamma}>1$ GeV,
in {\it a given direction $\theta$ with respect to the jet axis}.

Depending on the characteristics of the primary star and the geometry of the system, high-energy gamma rays can be absorbed in the anisotropic stellar photon field through pair production, and then inverse Compton emission from these pairs can initiate an $e^{\pm}$-pair cascade. This effect has been studied in detail by Bednarek (1997). The main effect of these cascades is a degradation of TeV gamma-rays into a form of softer MeV-GeV emission. Very close systems ($a\sim 10^{11}$ cm) with O stars and perpendicular jets can be optically thick for gamma-rays between $\sim 0.1-100$ TeV. If the jet is inclined towards the star, the effect can be stronger (Bednarek 1997). For systems with larger separations, the opacity rapidly falls below unity.

\begin{figure}
\resizebox{7.5cm}{!}{\includegraphics{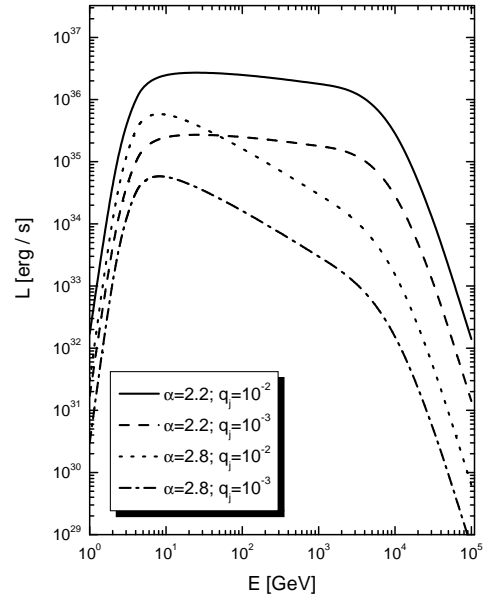}} \caption{\rm
Spectral high-energy distribution for windy MQs with
proton index  $\alpha=2.2$ and $\alpha=2.8$, for different
jet/disk coupling constants ($q_j$). The jet inclination with
respect to the line of sight is assumed to be 10 degrees. An angle
of 30 degrees reduces the luminosity in about two orders of
magnitude.} \label{fig2}
\end{figure}

In order to make some numerical estimates, we shall adopt the
specific MQ model presented in Table \ref{t1}. The values
chosen for the different parameters are typical for MQs
with O stellar companions, like Cygnus X-1. We shall consider a
conical jet ($\epsilon=1$) with conservation of the number of
protons ($n=2$) and a high-energy cutoff for the population of
relativistic protons of ${E'}_p^{\rm max}=100$ TeV. The minimum distance from the jet to the primary star is $a=70\;R_{\sun}\sim 5 \times 10^{12}$ cm.
\begin{table} 
\caption{Basic parameters of the model}
\begin{tabular}{lll}
\hline
Parameter & Symbol  & Value  \\
\hline
Type of jet &  $\epsilon$ & 1 \\
Black hole mass & $M_{\rm bh}$ & 10 $M_{\sun}$\\
Injection point & $z_0$ & 50 $R_{\rm g}^1$  \\
Initial radius & $R_0$ & 5 $R_{\rm g}$ \\
Radius of the companion star & $r_*$ &35 $R_{\sun}$ \\
Mass loss rate & $\dot{M}_*$ & $10^{-5}$ $M_{\sun}$ yr$^{-1}$ \\
Terminal wind velocity & $v_{\infty}$ & 2500 km s$^{-1}$\\
Black hole accretion rate & $\dot{M}_{\rm disk}$ & $10^{-8}$ $M_{\sun}$
yr$^{-1}$ \\
Wind velocity index & $\beta$ & 1 \\
Jet's expansion index & $n$ & 2 \\
Jet's Lorentz factor & $\Gamma$ & 5 \\
Minimum proton energy & ${E'}_p^{\rm min}$ & 10 GeV \\
Maximum proton energy & ${E'}_p^{\rm max}$ & 100 TeV \\
Orbital radius & $a$ & 2 $r_*$\\
\hline
\multicolumn{3}{l} {$^1$$R_{\rm g}=GM_{\rm bh}/c^2$.}\cr
\end{tabular}
\label{t1}
\end{table}
In Fig.~\ref{fig2} we show the spectral high-energy distribution
for models with proton index $\alpha=2.2$ and $\alpha=2.8$, and
for different values of the jet/disk coupling parameter $q_{\rm
j}$, with the results obtained using a numerical integration
routine.
We have added an exponential-like cutoff at $E_{\gamma}\sim 0.1
{E'}_p^{\rm max}$.
%
%
Since the forward momentum of the protons in the jet is so great,
the gamma-rays will be highly beamed, within an angle $\Theta\sim$
arctg $(R/z)$. Hence, only hadronic microblazars would be
detected.

\section{Discussion}

Neutrinos are also generated by pion production chains.
The signal-to-noise
(S/N) ratio for the detection of such a $\nu$-signal can be
obtained analyzing the event rate of atmospheric $\nu$-background
and comparing it with the event rate from the source (see e.g.
Anchordoqui et al. 2003 for details).
The S/N ratio in a km-scale detector (like ICECUBE) in the 1--10
TeV band (including the effects of neutrino oscillations) is $\sim
3$ for one year of operation, assuming an inclination angle of 30
degrees, $\alpha=2.2$ and that the neutrino spectrum roughly
satisfies (e.g. Dar \& Laor 1997): $
{dF_{\nu}}/{dE_{\nu}}\simeq0.7 {dF_{\gamma}}/{dE_{\gamma}}. $ This
S/N is high enough as to justify speculations on the possibility
of detecting a hadronic microblazar first from its neutrino signal
(a serendipitious discovery in a detector like ICECUBE), and only
later from its gamma-ray emission (through a pointed observation). Note that this neutrino flux is
different from, and in some cases can even be stronger than, the
neutrino flux produced by photomeson processes within the coronal
region (Levinson and Waxman 2001, Distefano et al. 2002), thus the detection of the
neutrino emission from MQs seems promising. It is not possible to separate in the neutrino signal the contributions form the photomeson and $p-p$ channels, but simultaneous X-ray observations can help to determine the characteristics of the relevant photon fields to which the inner jet is exposed, making then possible estimates of each contribution in particular cases.


Another interesting aspect of the hadronic microblazar is that
$e^{\pm}$ are injected outside the coronal region through
$\pi^{\pm}$ decays. These leptons do not experience the severe
IC losses that affect to primary electrons and pairs
(Romero et al. 2002). These secondaries will mainly cool through
synchrotron radiation (at X-rays, in the case of TeV particles)
and IC interactions with the stellar seed photons (that would
result into an additional source of MeV-GeV gamma-rays). The
spectrum of secondary pairs roughly mimics the shape of the proton
spectrum. Hence, synchrotron emission from these particles will
present indices $\alpha_{\rm syn}\simeq (\alpha-1)/2$, which for
values of $\alpha\sim2$ are similar to what is observed in
MQs' jets at radio wavelengths. The losses of the
primaries in the inner source lead to a soft particle spectrum that is then injected in
the region where the particles produce IC gamma-rays through
interactions with stellar UV photons.  Pure leptonic models for
the gamma-ray flux (e.g. Kaufman-Bernad\'o et al. 2002) then
require particle re-acceleration in order to explain the flat
radio-spectrum of sources like LS 5039 far from the core. In the
hadronic model, the leptons are injected {\em in situ} with the
right spectrum through hadronic decays, avoiding the problem.

At TeV gamma-ray energies hadronic microblazars which are optically thin to pair production can be detected as
unidentified, point-like sources with relatively hard spectra.
This kind of sources could display variability.
In the near future, new ground-based \v{C}erenkov telescopes like HESS
and MAGIC might detect the signatures of such sources on the
galactic plane. Hadronic microblazars might be part of this
population, as well as of the parent population of low latitude
unidentified EGRET sources.

\begin{acknowledgements}

G.E.R. is very grateful to F. Aharonian, H. V\"olk and all the
staff of the Max-Planck-Institut f\"ur Kernphysik at Heidelberg
for their kind hospitality during the development of this
research. We thank
valuable comments by F. Aharonian, P.L. Biermann, L. Costamante, I. Grenier, C-Y. Huang, D. Purmohammad, and
M. Rib\'o. G.E.R. is supported by Fundaci\'on
Antorchas (also M.M.K.B.), ANPCyT (PICT 03-04881), and CONICET
(PIP 0438/98). D.F.T. research is done under the auspices of the
US Department of Energy (NNSA), by the UC's LLNL under contract
W-7405-Eng-48. This research benefited from the ECOS
French-Argentinian cooperation agreement. We thank an anonymous referee for valuable comments.

\end{acknowledgements}

{}

\end{document}